\documentclass[12pt,preprint]{aastex}

\usepackage{fullpage}
\usepackage{bm}		%

\usepackage{booktabs}
\usepackage{latexsym}
\usepackage{amsmath}
\usepackage{amssymb}
\usepackage{ascmac}
\usepackage{graphicx}
\usepackage{theorem}
\usepackage{color}
\usepackage{subfigure}
\usepackage{float}

\newcommand{\bfm}[1]{\mbox{\boldmath{$#1$}}}

\shortauthors{Hirabayashi}

\begin{document}


\title{Structural failure of two-density-layer cohesionless biaxial ellipsoids}


\author{Masatoshi Hirabayashi}
\affil{Aerospace Engineering Sciences, 429 UCB, University of Colorado, Boulder, CO 80309-5004 United States}
\email{masatoshi.hirabayashi@colorado.edu}





\begin{abstract}
This paper quantitatively evaluates structural failure of biaxial cohesionless ellipsoids that have a two-density-layer distribution. The internal density layer is modeled as a sphere, while the external density layer is the rest of the part. The density is supposed to be constant in each layer. The present study derives averaged stresses over the whole volume of these bodies and uses limit analysis to determine their global failure. The upper bound condition of global failure is considered in terms of the size of the internal layer and the aspect ratio of the shape. The result shows that the two-density-layer causes the body to have different strength against structural failure.
\end{abstract}





\section{Introduction}
Theoretical studies about plastic analysis, using perfect ellipsoids, have been recently successful in explaining deformation processes of small bodies. \cite{Holsapple2001} firstly used limit analysis to calculate the limit spins of cohesionless rubble pile ellipsoids and showed that any elements in a perfect ellipsoid reach the limit state at the same time. \cite{Holsapple2004} reported that the limit spin given by volume averaged stresses is identical to the analysis given by \cite{Holsapple2001}. Using the limit analysis approach, \cite{Holsapple2006} investigated the tidal disruption condition of cohesionless ellipsoids. \cite{Sharma2009}, on the other hand, introduced an averaging form over the whole volume of dynamical deformation to discuss structural failure of a rubble pile ellipsoid due to a tidal effect of a massive body. \cite{Sharma2010} extended the theory by \cite{Sharma2009} to binary-ellipsoid systems. 

Regardless of their elegant mathematical formulations, these researches simplified their discussion by ignoring the effect of density distribution. It is interesting to understand how the density distribution has an effect on structural stability. In asteroid environments, the density may be distributed axisymmetrically during its accretion process. The present study considers the effect of this axisymmetric density distribution on the structural stability of asteroids. To investigate this effect, a uniformly rotating asteroid is modeled as a biaxial ellipsoid composed of an internal sphere and an external shell. The density is assumed to be constant in each layer. The technique used here is based on the limit analysis technique by \cite{Holsapple2004} who obtained the limit spin by using the total volume stress. The current paper is organized as follows. First, the two-layer model is established. Second, the limit analysis technique is applied to the two-layer model. Last, the upper bound condition of structural failure of the present model is compared with that of the uniform density case. 

\section{Two-layer model of a rubble pile biaxial ellipsoid}
\subsection{Definition}
\label{sec:def}
A biaxial ellipsoid with dimensions of $2 a$ by $2 b$ by $2 b$, where $a>b$, is supposed to be spinning with a constant spin $\omega$ along the maximum principal axis. This biaxial ellipsoid is composed of an internal sphere with a radius $l b$, where $l$ is less than $1$, and an external layer enclosing the internal layer. The volume of the internal layer can vary as $l$ changes. Those layers are concentric. Let us denote the density of the internal layer by $\rho$, the density of the external layer by $\rho^\prime$, and the averaged density of the whole volume by $\rho^\ast$. 

Since the following discussion will use normalized forms, definitions of mathematical notations are given first. Any lengths are normalized by $a$, and the size of a two-layer biaxial ellipsoid is characterized by the aspect ratio $\beta=b/a$. Similarly, a normalized position of an arbitrary element is denoted by $(x_1,x_2,x_3)$. The dimensionless spin rate is defined by $\Omega = \omega/\sqrt{\pi \rho^\ast G}$, where G is the gravitational constant. The density is normalized by the averaged density; in the formulation, a scale density relative to the averaged density, denoted as $\epsilon$, will be used. In other words, $\rho/\rho^\ast = 1 + \epsilon$ and $\rho^\prime/\rho^\ast = 1 + \epsilon^\prime$. This paper considers the total mass to be constant, there is the relation between $\epsilon$ and $\epsilon^\prime$: 
\begin{eqnarray}
\epsilon^\prime = - \frac{\epsilon \beta l^3}{1-\beta l^3}. \label{Eq:ep}
\end{eqnarray}
Potential $U$, body force $\bfm b$, and stress tensor $\bfm T$ are normalized by $\pi \rho^\ast G a^2$, $\pi \rho^\ast G a$, and $\pi \rho^{\ast 2} G a^2$, respectively. For $\bfm b$ and $\bfm T$, the following discussions will use index notations instead of vector notations. With indices $(i,j) = (1,2,3)$, these vector notations are expressed by $b_i$ and $T_{ij}$, respectively. 

The outer boundary of the internal layer and that of the external layer are introduced. The outer boundary of the internal layer is given as 
\begin{eqnarray}
x_1^2 + x_2^2 + x_3^2 = l^2 \beta^2.
\end{eqnarray}
On the other hand, the outer boundary of the external layer, or the surface of a biaxial body, is given as 
\begin{eqnarray}
x_1^2 + \frac{x_2^2 + x_3^2}{\beta^2} = 1. 
\end{eqnarray}
The stress state of this problem is given by the equilibrium equation: 
\begin{eqnarray}
\frac{\partial T_{ij}}{\partial x_i} + (1 + \epsilon_k) b_j = 0, \label{Eq:Eqn_Eqb}
\end{eqnarray}
where $\epsilon_k = \epsilon$ if an element is in the internal layer and $\epsilon_k = \epsilon^\prime$ if an element is in the external layer. 

\subsection{Calculation of body forces}
This study focuses on the effect of a gravitational force and a centrifugal force, so $b_i$ is a function of the density and the spin rate. The acceleration components of a centrifugal force $b_{c,i}$ are given as
\begin{eqnarray}
b_{c,i} = \left\{ 
\begin{array}{l l}
\Omega^2 x_i, & \text{if $\quad i=1,2$} \\
0, & \text{if $\quad i = 3$}
\end{array} \right.
\end{eqnarray}
On the other hand, the gravity computation requires considerations of density distribution. Here, a combination of a uniform-density ellipsoid and a uniform-density sphere is considered to obtain a gravitational acceleration. The potential can be described as
\begin{eqnarray}
U &=& - \frac{1}{\pi} \int_{V} \frac{1+\epsilon(\bfm r)}{d} dV, \nonumber \\
&=& - \frac{1}{\pi} \int_{V_{ex}} \frac{1+\epsilon^\prime}{d} dV - \frac{1}{\pi} \int_{V_{in}} \frac{1+\epsilon}{d} dV, \nonumber \\
&=& - \frac{1}{\pi} \int_{V} \frac{1+\epsilon^\prime}{d} dV - \frac{1}{\pi} \int_{V_{in}} \frac{\epsilon-\epsilon^\prime}{d} dV, \label{Eq:Potential}
\end{eqnarray}
where $\epsilon (\bfm r)$ is the scale density at an arbitrary element, $V$ is the total volume, $V_{ex}$ is the volume of the external layer, $V_{in}$ is the volume of the internal layer, and $d$ is the distance between two small elements. The third row indicates that computation of a gravitational acceleration can be decoupled into a perfect ellipsoid and a perfect sphere. The first term in the third row in Eq. (\ref{Eq:Potential}), denoted as $U_{el}$, is written as 
\begin{eqnarray}
U_{el}  = - (1+\epsilon^\prime) (A_0 + \sum_{i=1}^3 A_i x_i^2), 
\end{eqnarray}
where 
\begin{eqnarray}
A_0 &=& \beta^2 \int_0^\infty \frac{d s}{(s+ \beta^2) \Delta}, \\
A_1 &=& \beta^2 \int_0^\infty \frac{d s}{(s+1) (s+ \beta^2) \Delta}, \label{Eq:Ax} \\
A_2 &=& A_3 = \beta^2 \int_0^\infty \frac{d s}{(s+\beta^2)^2 \Delta}. \label{Eq:Az} 
\end{eqnarray}
and $\Delta = \sqrt{s + 1}$. The second term in the third row in Eq. (\ref{Eq:Potential}), denoted as $U_{sp}$, is given as 
\begin{eqnarray}
U_{sp} = 
\left\{ 
	\begin{array}{l l}
    		- \frac{4 l^3 \beta^3 (\epsilon-\epsilon^\prime)}{3r}  & \quad \text{if $r > l \beta$}, \\
    		- \frac{4 r^2 (\epsilon-\epsilon^\prime)}{3} & \quad \text{if $r \le l \beta$},
 	\end{array} 
\right.
\end{eqnarray}
where $r$ is the distance between a field point and the center of mass. 

Differentiating those potentials with respect to the position yields the gravitational acceleration:
\begin{eqnarray}
b_{g,i} = 
\left\{ 
	\begin{array}{l l}
    		- 2 A_i (1 + \epsilon^\prime) x_i - \frac{4 l^3 \beta^3 (\epsilon-\epsilon^\prime)}{3} \frac{x_i}{r^3}, & \quad \text{if $r > l \beta$}, \\
    		- 2 A_i (1 + \epsilon^\prime) x_i - \frac{4 (\epsilon-\epsilon^\prime)}{3} x_i. & \quad \text{if $r \le l \beta$},
 	\end{array} 
\right.
\end{eqnarray}
The first row indicates the gravitational acceleration in the external layer, while the second row describes that in the internal layer. The total body forces, a sum of the centrifugal and gravitational accelerations $b_i = b_{gi} + b_{ci}$, are given as follows. The body force in the external layer $b_{ex,i}$ is given as
\begin{eqnarray}
b_{ex,i} = - 2 A_i (1 + \epsilon^\prime) x_i - \frac{4 l^3 \beta^3 (\epsilon - \epsilon^\prime)}{3} \frac{x_i}{r^3} + \Omega_i^2 x_i.
\end{eqnarray}
On the other hand, the body force in the internal layer $b_{in,i}$ is obtained as
\begin{eqnarray}
b_{in,i} = 
- 2 A_i (1+\epsilon^\prime) x_i - \frac{4 (\epsilon - \epsilon^\prime)}{3} x_i + \Omega_i^2 x_i.  
\end{eqnarray}
Note that $[\Omega_1, \Omega_2, \Omega_3] = [\Omega, \Omega, 0]$. 

\section{Upper bound condition of structural failure}
This study assumes that materials are characterized by elastic perfectly-plastic theory, a smooth-convex yield envelope, and an associate flow. In limit analysis, the upper bound theorem provides the condition where a target body must fail plastically (see \citealt{Chen1988}). \cite{Holsapple2008A} derived that the upper bound condition is identical to the yield condition of averaged stresses over an arbitrary volume. The present paper utilizes this technique to determine the upper bound of structural failure of the whole volume. The yield condition is modeled by using the Mohr-Coulomb yield criterion, which is given as
\begin{eqnarray}
g (\sigma_1, \sigma_3, \phi) \le 0, \label{Eq:MC}
\end{eqnarray}
where 
\begin{eqnarray}
g (\sigma_1, \sigma_3, \phi) = \frac{\sigma_1 - \sigma_3}{2} \sec \phi + \frac{\sigma_1 + \sigma_3}{2} \tan \phi,
\end{eqnarray}
and $\phi$ is the angle of internal friction. If materials are cohesive, the term for cohesive strength should appear in the right hand side in Eq. (\ref{Eq:MC}). It is necessary to clarify the use of this yield criterion, which is not smooth at a compression meridian and a tension meridian, for limit analysis. A biaxial ellipsoid spinning along the maximal principal axis experiences the stress states at the meridians in some conditions (see Fig. 3 of \citealt{Holsapple2001}). However, since this condition may be unrealistic and limited in nature, this paper does not consider such a condition. Therefore, since stress states are always between these meridians and the yield envelope is smooth in this region, the Mohr-Coulomb yield envelope is still applicable to the current problem. Note that the use of the Drucker-Prager yield criterion removes this assumption. Nevertheless, the technical method used here does not change due to yield conditions, so this paper uses the Mohr-Coulomb yield criterion.

There is a standard formula for the total volume stress. Using the general form yields the total volume stress of a two-layer biaxial ellipsoid:
\begin{eqnarray}
\bar T^t_{ij} = \frac{1}{V_t} \int_{V_{ex}} x_j b_{ex,i} dV_{ex} + \frac{1}{V_t} \int_{V_{in}} x_j b_{in,i} dV_{in}. \label{Eq:Total1}
\end{eqnarray}
Since the diagonal components of the stress tensor are zero, Eq. (\ref{Eq:Total1}) can be simply rewritten as  
\begin{eqnarray}
\bar T^t_{ii} &=& - \frac{(1 + \epsilon^\prime) [2 A_i (1 + \epsilon^\prime)  - \Omega_i^2]}{V} E_{1,i}  \nonumber \\
&& - \frac{4 l^3 \beta^3 (1+\epsilon^\prime) (\epsilon-\epsilon^\prime)}{3 V} E_{2,i} \nonumber \\
&& - \frac{(1 + \epsilon) [2 A_i (1+\epsilon^\prime)  + 4 (\epsilon-\epsilon^\prime)/3 - \Omega_i^2]}{V} F_i, \label{Eq:totalStress}
\end{eqnarray}
where $V = 4 \pi \beta^2/3$ and $[\Omega_1, \Omega_2, \Omega_3] = [\Omega, \Omega, 0]$. $E_{1,i}$, $E_{1,i}$, and $F_i$ are given as
\begin{eqnarray}
E_{1,i} &=& \int_{V_{ex}} x_i^2 dV_{ex}, \nonumber \\
E_{2,i} &=& \int_{V_{ex}} \frac{x_i^2}{d^3} dV_{ex}, \nonumber \\
F_i &=& \int_{V_{in}} x_i^2 dV_{in}. \nonumber 
\end{eqnarray}

Finally, substitution of Eq. (\ref{Eq:totalStress}) into the yield condition $g(\sigma_1, \sigma_3, \phi) = 0$ gives the upper bound condition of structural failure.

\section{Application to small bodies}
This section considers comparison of the upper bound condition of the present model and that of the uniform density case. 

\begin{figure}[ht!]
	\begin{center}
		\subfigure[]{
         		\label{Fig:denDisA}	
		\includegraphics[width=0.45\textwidth]{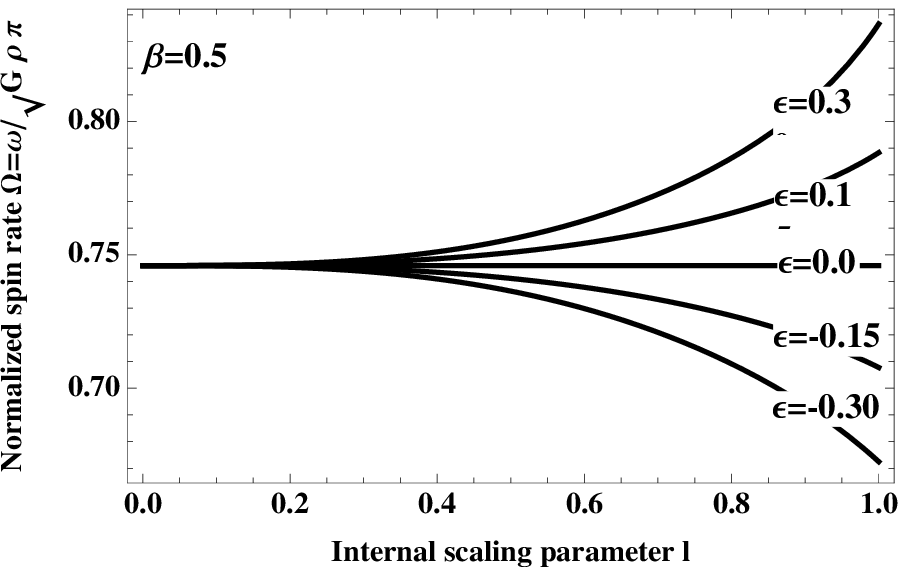}
          	}
		\subfigure[]{
         		\label{Fig:denDisB}	
		\includegraphics[width=0.45\textwidth]{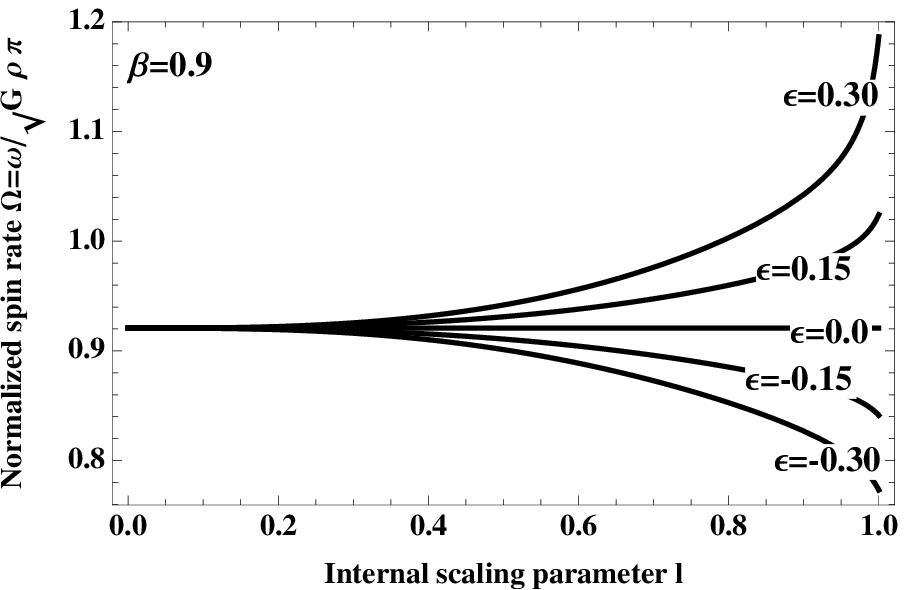}
          	}
	\caption{Change of the critical spin rate with regards to the scaling parameter $l$. Since the mass is constant, for each $\epsilon$, $\epsilon^\prime$ is calculated by using Eq. (\ref{Eq:ep}). Fig. \ref{Fig:denDisA} shows the case $\beta=0.5$, while Fig. \ref{Fig:denDisB} gives the case $\beta=0.9$. The friction angle is chosen as 30$^\circ$.}
	\label{Fig:denDis}
	\end{center}
\end{figure}

First, the effect of the scaling parameter on the critical spin $\Omega$ is discussed. Figure \ref{Fig:denDis} shows change of the critical spin with regards to the scaling parameter $l$. The friction angle is chosen as $30^\circ$. Fig. \ref{Fig:denDisA} gives the case $\beta=0.5$, while Fig. \ref{Fig:denDisB} describes the case $\beta=0.9$, where $\beta$ is the aspect ratio. Normalization defined above allows for the mass constant condition; in other words, the mass is constant in each plot. The case $\epsilon=0.0$ in both plots is consistent with the uniform density case (e.g., \citealt{Holsapple2001}). It can be found that if $\epsilon>0 (<0)$, i.e., high (low) density in the internal layer, the critical spin rate increases (decreases) as $l$ becomes larger. Therefore, the body becomes stronger (weaker) against structural failure if $\epsilon>0 (<0)$. Differences between the case $\beta=0.5$ and $\beta=0.9$ can also be seen. Compared to the case $\beta=0.5$, the slope of the critical spin for the case $\beta=0.9$ becomes steeper as $l$ increases.  

Second, the density distribution case is compared with the uniform density case in terms of the aspect ratio\footnote{The latter case corresponds to Fig. 8 by \cite{Holsapple2001}.}. Here, the asteroid LightCurve Data Base (LCDB) by Warner, Harris, and Pravec (revised on November 10, 2012) is also given. Instructed by its manual, the following discussion only uses the objects of which $U$ (quality) code is more than or equal to 2. In addition, since the spin barrier is split into the gravity regime and the cohesive regime (\citealt{Holsapple2007}), only asteroids  in the gravity regime ranging between 5 km and 300 km are considered. The LCDB data includes the spin periods, sizes, and observational full-range amplitudes. This study uses a standard formula for the largest observed amplitude relative to the aspect ratio:
\begin{eqnarray}
A = - 2.5 \log \beta,
\end{eqnarray}
where $A$ is the observational amplitude. Again, the smallest diameter is assumed to be equal to the intermediate diameter here. Also, the averaged density is fixes as 2.5 g/cm$^3$.

\begin{figure}[p!]
\begin{center}
\includegraphics[width=5in]{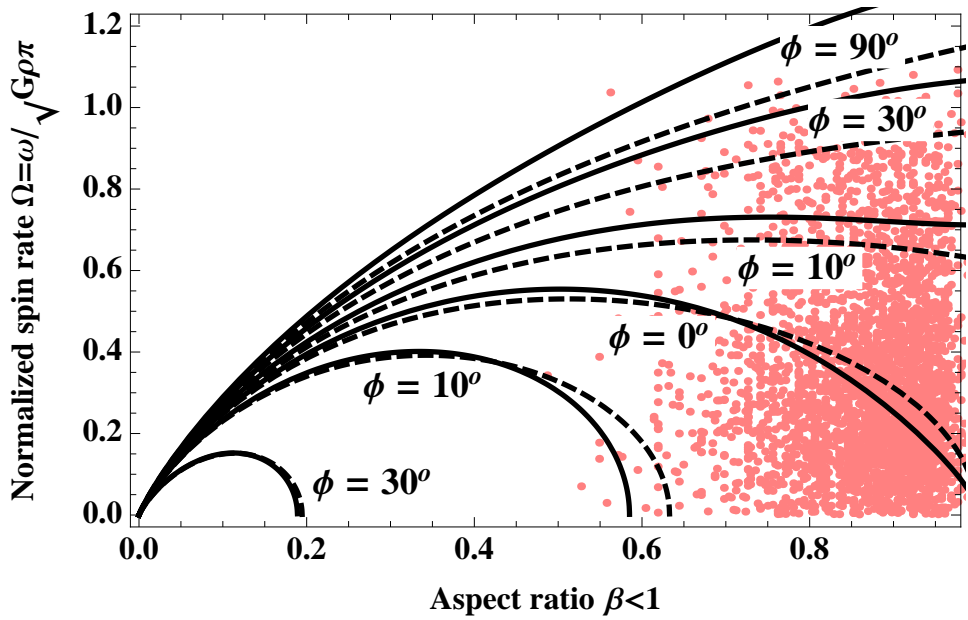}
\caption{Critical spin rate in terms of the aspect ratio. The solid lines indicate the critical spin rate for the case $l=0.9$ and $\epsilon=0.3$, while the dashed lines describe the uniform density case, which is consistent with Fig. 8 by \cite{Holsapple2001}. For the LCDB data plotted here, the averaged density is assumed to be 2.5 g/cm$^3$.}
\label{Fig:denDisC}
\end{center}
\end{figure}

Figure \ref{Fig:denDisC} gives the critical spin rate with regards to the aspect ratio. The solid lines show the critical spin for the density distribution case, while the dashed lines give that for the uniform density case. Each line is calculated based on a different friction angle. For the density distribution case, the external and internal layers are characterized by $\epsilon = 0.3$ and $l=0.9$. This condition gives a high density core and a low density surface. It is found that the critical spins become higher in the tension regime. On the other hand, interestingly, in the compression regime, the critical spins for the density distribution case is higher than those for the uniform density case when $\beta$ is small\footnote{In the compression regime, the critical spins are the minimum spins that suspend the bodies.}.

\section{Discussion and Conclusion}
This paper investigated the effect of the two-layer density distribution on structural failure of a uniformly rotating ellipsoid. The prime result shows that the two-layer density distribution causes different failure conditions from a uniformly rotating ellipsoid. The larger (smaller) size and higher (lower) density of the internal layer allow the bodies to become stronger (weaker) against structural failure. On the other hand, the critical spins with regards to the aspect ratio behave differently. If there is a high density core, the critical spins in the tension regime can increase, which indicates that the body becomes stronger against tension failure. However, in the compression regime, the critical spins for the density distribution case is  also larger than those for the uniform density case if $\beta$ decreases. This implies that the body becomes weaker against compression failure. It can be explained that for such elongated bodies, since a dense core causes stronger gravitational compression and centrifugal forces do not effectively support the bodies, the bodies become sensitive to structural failure. 



\acknowledgments
The author wishes to thank Dr. Holsapple for his detailed reviews that improved the clarity and quality of the manuscript.

\bibliographystyle{model2-names}
\bibliography{refInterior}  




\end{document}